\documentclass {elsart}
\usepackage[english]{babel}
\usepackage {amssymb,graphicx,subfigure}
\usepackage{cite}
\makeatletter
\def\anti#1{\mathpalette{\@anti}{#1}#1}%               % better "anti" overbar
\def\@anti#1#2{\sbox0{$#1#2$}%                         % auxiliary for .......
  \makebox[0pt][l]
    {$#1\kern.30\ht0\overline{\kern-.35\ht0\phantom{#2}\kern-.1ex}$}}
\makeatother
\def\spig{\kern.3em\raise1.2ex\hbox{$|$}\kern-.48em\to}% bent right arrow
\def\http://{http:/\kern-.2em/}

\makeatletter
\renewcommand\section{\@startsection {section}{1}{\z@}%
                                   {-3.5ex \@plus -1ex \@minus -.2ex}%
                                   {2.3ex \@plus.2ex}%
                                   {\normalfont\bfseries}}
\renewcommand\subsection{\@startsection{subsection}{2}{\z@}%
                                     {-3.25ex\@plus -1ex \@minus -.2ex}%
                                     {1.5ex \@plus .2ex}%
                                     {\normalfont\bfseries}}
\makeatother
\newcommand\scatt{\mathrm{scatt}}
\def\bibsubsection#1{\vspace{3ex}\hbox to\textwidth{\bf#1\hfill}\smallskip}
\def\bibsubsubsection#1{\hbox to\textwidth{\bf#1\hfill}\smallskip}

\begin{document}
\begin{frontmatter}

\title{\boldmath
  Dalitz plot analysis of $D_s^+$ and $D^+$ decay to $\pi^+\pi^-\pi^+$ using
  the \emph{K}-matrix formalism.
}

\author[ucd]{J.~M.~Link}
\author[ucd]{P.~M.~Yager}
\author[cbpf]{J.~C.~Anjos}
\author[cbpf]{I.~Bediaga}
\author[cbpf]{C.~G\"obel}
\author[cbpf]{A.~A.~Machado}
\author[cbpf]{J.~Magnin}
\author[cbpf]{A.~Massafferri}
\author[cbpf]{J.~M.~de~Miranda}
\author[cbpf]{I.~M.~Pepe}
% ...

\address[ucd]{University of California, Davis, CA 95616}
\address[cbpf]{Centro Brasileiro de Pesquisas F\'\i sicas, Rio de Janeiro, RJ, Brasil}
\address[cinv]{CINVESTAV, 07000 M\'exico City, DF, Mexico}
\address[cu]{University of Colorado, Boulder, CO 80309}
\address[fnal]{Fermi National Accelerator Laboratory, Batavia, IL 60510}
\address[fras]{Laboratori Nazionali di Frascati dell'INFN, Frascati, Italy I-00044}
\address[guan]{University of Guanajuato, 37150 Leon, Guanajuato, Mexico}
\address[ui]{University of Illinois, Urbana-Champaign, IL 61801}
\address[iu]{Indiana University, Bloomington, IN 47405}
\address[korea]{Korea University, Seoul, Korea 136-701}
\address[korea2]{Kyungpook National University, Taegu, Korea 702-701}
\address[milan]{INFN and University of Milano, Milano, Italy}
\address[nc]{University of North Carolina, Asheville, NC 28804}
\address[pavia]{Dipartimento di Fisica Nucleare e Teorica and INFN, Pavia, Italy}
\address[pr]{University of Puerto Rico, Mayaguez, PR 00681}
\address[sc]{University of South Carolina, Columbia, SC 29208}
\address[ut]{University of Tennessee, Knoxville, TN 37996}
\address[vu]{Vanderbilt University, Nashville, TN 37235}
\address[wisc]{University of Wisconsin, Madison, WI 53706}

%\endnote{\small E-mail address sandra.malvezzi@mi.infn.it (S.Malvezzi)}
\endnote{\small
  See http://www-focus.fnal.gov/authors.html for additional author information.
}

\begin{abstract}
  FOCUS results from Dalitz-plot analyses of $D^+_s$ and $D^+$ to
  $\pi^+\pi^-\pi^+$ are presented. The \emph{K-matrix} formalism is applied to
  charm decays for the first time, which allows us to fully exploit the already
  existing knowledge coming from light-meson spectroscopy experiments. In
  particular all the measured dynamics of the $S$-wave $\pi\pi$ scattering,
  characterized by broad/overlapping resonances and large non-resonant
  background, can be properly included. This paper studies the extent to which
  the \emph{K-matrix} approach is able to reproduce the observed Dalitz plot and
  thus help us to understand the underlying dynamics. The results are discussed
  along with their possible implications for the controversial $\sigma$ meson.
\end{abstract}


\end{frontmatter}

\section{Introduction}
Charm-meson decay dynamics has been extensively studied in the last decade. The
analysis of the three-body final state by fitting Dalitz plots has proved to be
a powerful tool for investigating effects of resonant substructure,
interference patterns, and final state interactions in the charm sector. The
isobar formalism, which has traditionally been applied to charm amplitude
analyses, represents the decay amplitude as a sum of relativistic Breit-Wigner
propagators multiplied by form factors plus a term describing the angular
distribution of the two body decay of each intermediate state of a given spin.
Many amplitude analyses require detailed knowledge of the light-meson sector.
In particular, the need to model intermediate scalar particles contributing to
the charm meson in the decays reported here has caused us to question the
validity of the Breit-Wigner approximation for the description of the relevant
scalar resonances \cite{pdg_rev,penn1}. Resonances are associated with poles of
the \emph{S-matrix} in the complex energy plane. The position of the pole in
the complex energy plane provides the fundamental, model-independent,
process-independent resonance description. A simple Breit-Wigner amplitude
corresponds to the most elementary type of extrapolation from the physical
region to an unphysical-sheet pole. In the case of a narrow, isolated
resonance, there is a close connection between the position of the pole on the
unphysical sheet and the peak we observe in experiments at real values of the
energy. However, when a resonance is broad and overlaps with other resonances,
then this connection is lost.  The Breit-Wigner parameters measured on the real
axis (mass and width) can be connected to the pole-positions in the complex
energy plane only through models of analytic continuation.

A formalism for studying overlapping and many channel resonances has been
proposed long ago and is based on the  \emph{K-matrix} \cite{wigner,chung}
parametrization.
This formalism, originating in the context of two-body
scattering, can be generalized to cover the case of production of resonances in
more complex reactions \cite{aitch}, with the assumption that the two-body system in the
final state is an isolated one and that the two particles do not
simultaneously interact with the rest of the final state in the production
process \cite{chung}. The  \emph{K-matrix} approach allows us
to include the positions of the poles in the complex plane directly in our
analysis, thus directly incorporating the results from spectroscopy experiments
\cite{penn2,anisar1}. In addition, the \emph{K-matrix} formalism provides a
direct way of imposing the two-body unitarity constraint which is not
explicitly guaranteed in the simple isobar model. Minor unitarity violations
are expected for narrow, isolated resonances but more severe ones exist for
broad, overlapping states. The validity of the assumed quasi two-body nature of
the process of the \emph{K-matrix} approach can only be verified by a direct
comparison of the model predictions with data. In particular, the failure to
reproduce three-body-decay features would be a strong indication of the
presence of the neglected three-body effects.

\section{Candidate selection}

\begin{figure}[htb]
  \begin{center}
 \subfigure[$D_s^+$]
  {
  \includegraphics[angle=-90,width=0.4\textwidth]{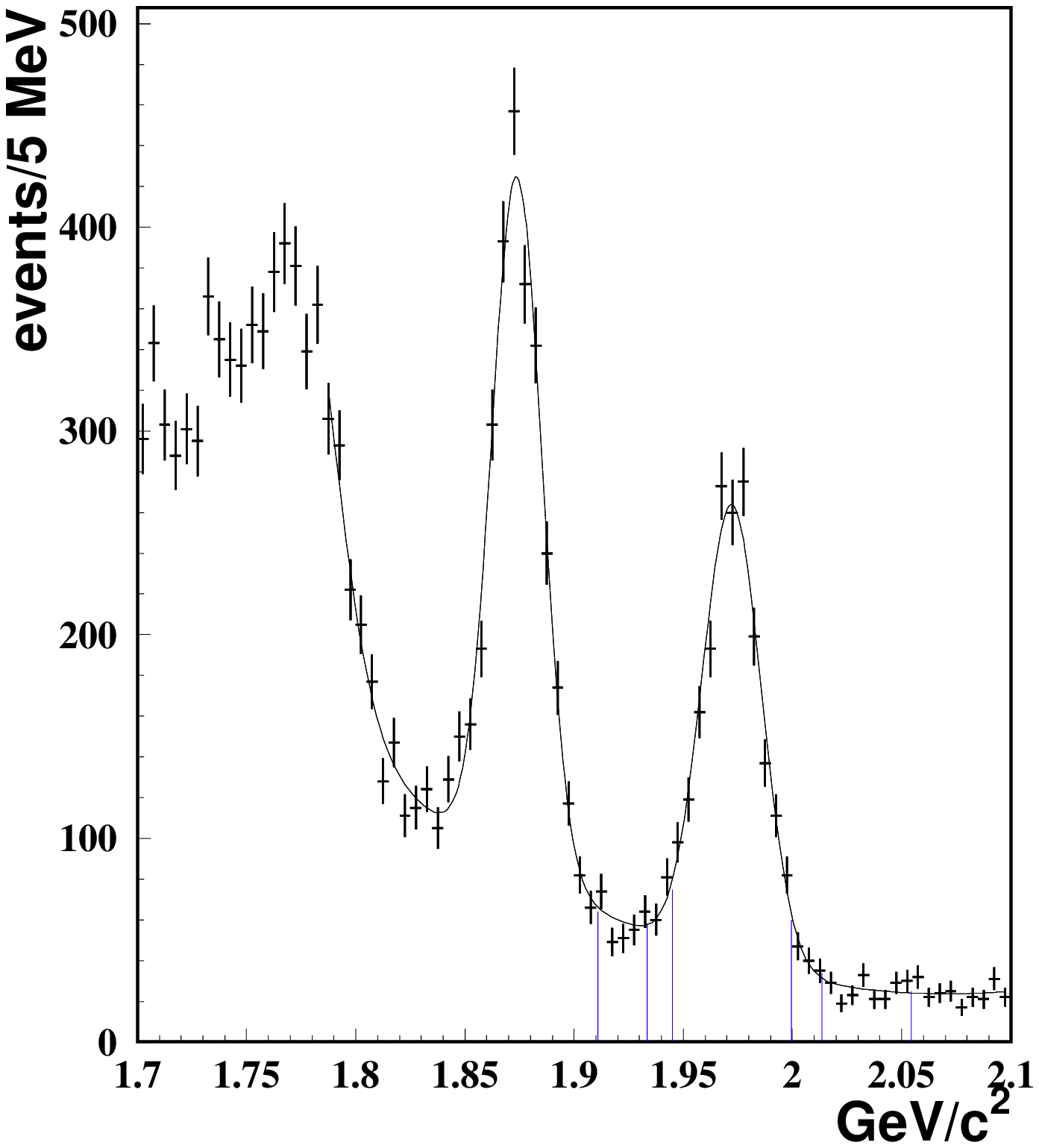}
  }
 \subfigure[$D^+$]
  {
  \includegraphics[angle=-90,width=0.4\textwidth]{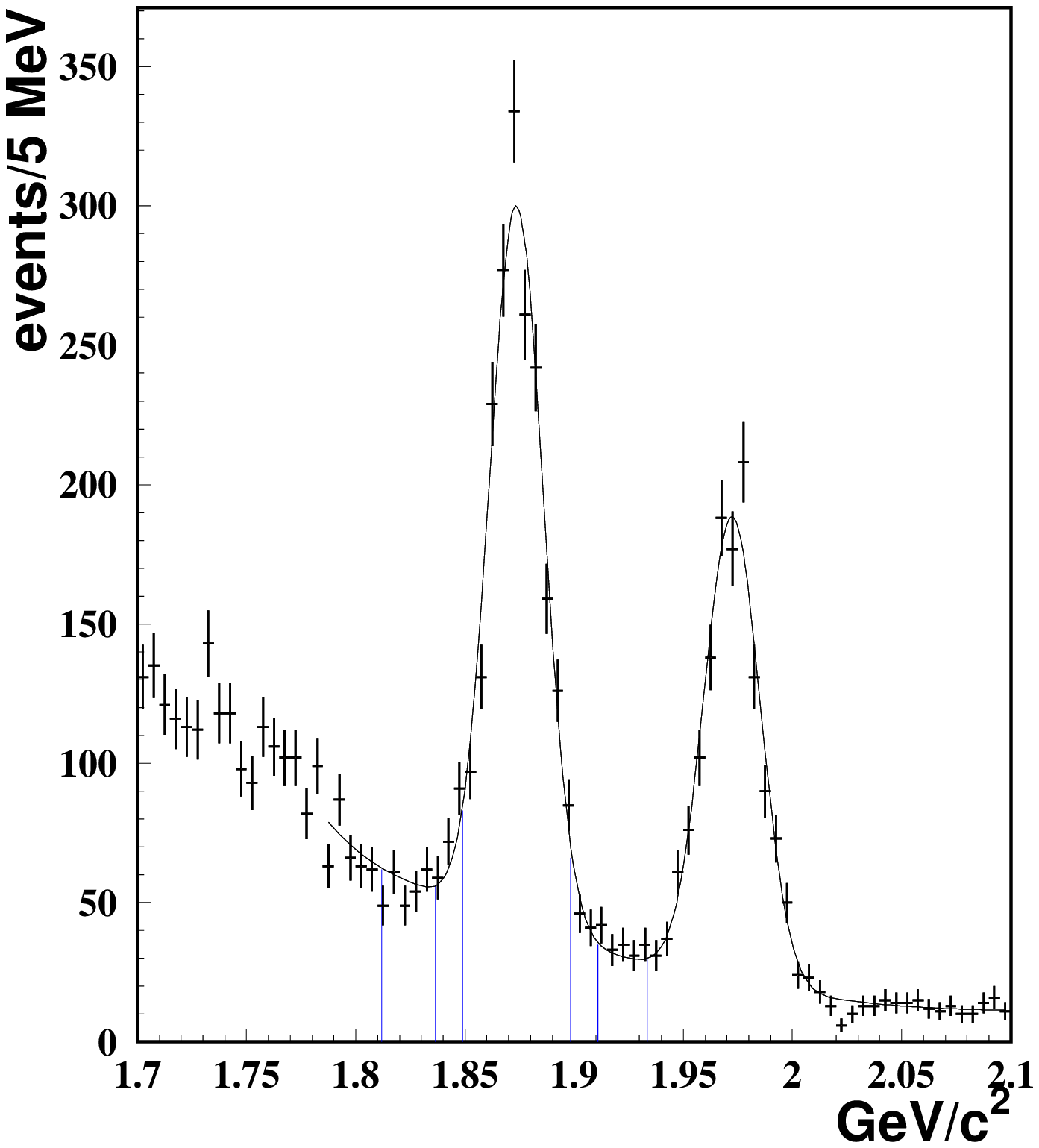}
  }
 \end{center}
  \caption{Signal and side-band regions of the three-pion invariant-mass distribution
  for a) $D_s^+$ and b) $D^+$ Dalitz-plot analysis respectively.}
  \label{mass}
\end{figure}

\begin{figure}[htb]
  \begin{center}
  \subfigure[$D_s^+$]
  {
  \includegraphics[width=0.47\textwidth]{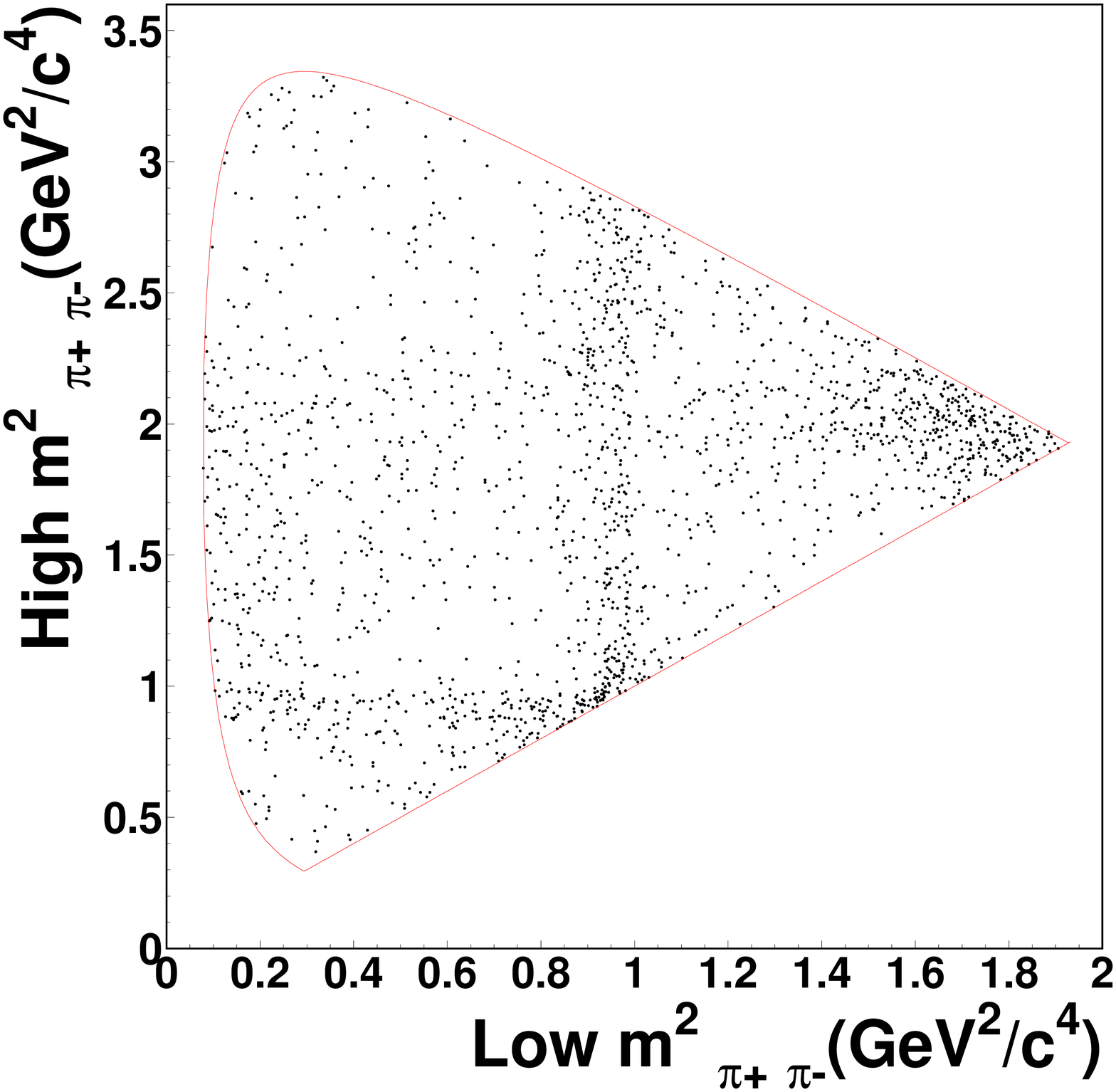}
  }
  \subfigure[$D^+$]
  {
  \includegraphics[width=0.47\textwidth]{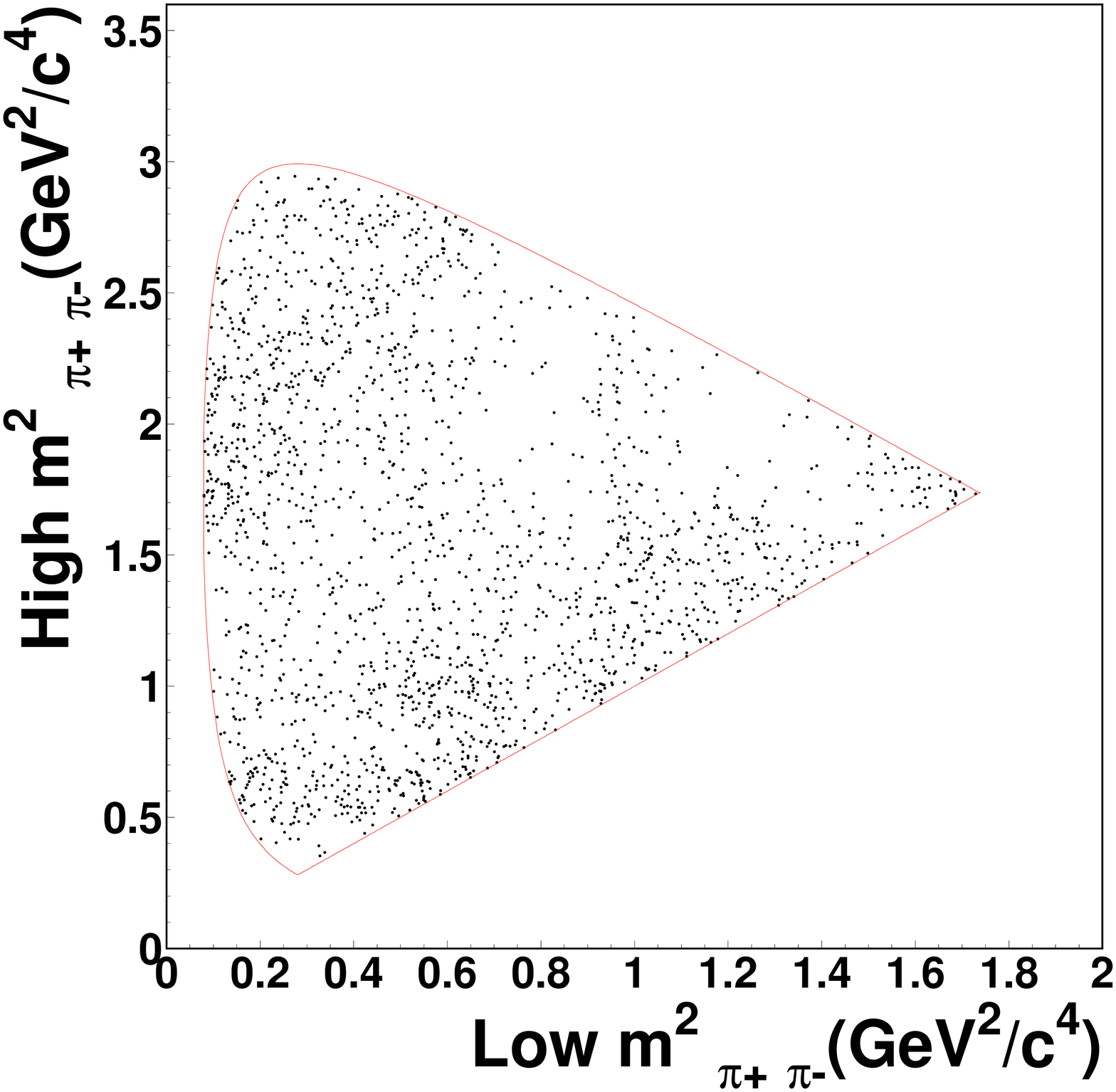}
  }
  \end{center}
  \caption{ a) $D_s^+$ and b) $D^+$ Dalitz plots.}
  \label{dalitz}
\end{figure}

The FOCUS detector is a large aperture, fixed-target spectrometer with
excellent vertexing and particle identification capabilities. We have chosen
cuts designed to minimize non-charm background as well as reflection
backgrounds from misidentified charm decays. The three-pion final states are
obtained using a candidate driven vertex algorithm. A decay vertex is formed
from three reconstructed charged tracks. The momentum of the $D$ candidate is
used to intersect other reconstructed tracks to form a production vertex. The
confidence levels (C.L.) of each vertex is required to exceed 1\,\%. After the
vertex finder algorithm, the variable $\ell$, which is the separation of the
primary and secondary vertex, and its associated error $\sigma_\ell$ are
calculated. We reduce backgrounds by requiring $\ell/\sigma_\ell>10$ and 7 for
the $D^+$ and $D_s^+$, respectively. The two vertices are also required to
satisfy isolation conditions. The primary vertex isolation cut requires that a
track assigned to the decay vertex has  a C.L. less than 1\,\% to be included
in the primary vertex. The secondary vertex isolation cut requires that all
remaining tracks not assigned to the primary and secondary vertex have a C.L.\
smaller than 0.1\,\% to form a vertex with the $D$ candidate daughters. The
decay vertex is required to be  $3\,\sigma$ outside of the target material to
reduce the background due to hadronic re-interactions in the material. A cut on
the negative log likelihood of the \v{C}erenkov hypothesis \cite{nimcer}  of
$W_K-W_{\pi}>0$ is required for each pion. A tighter cut of $W_K-W_{\pi}>5$ is
required on opposite-sign pion in the $D^+$ decay in order to remove the
$K^-\pi^+\pi^+$ reflection contribution to the $D^+$ low-mass sideband. We
further require that all three pions satisfy a loose pion-consistency cut of
$\Delta W=W_{\pi}-W_{\min}<3$ where $W_{\min}$ is the negative log likelihood
of the most favored \v{C}erenkov hypothesis. The vertex isolation requirement
nearly eliminates $D^{*+}\to D^0\pi^+$ contamination. The samples selected
according to these requirements (Fig.~\ref{mass}) consist of $1475\pm50$ and
$1527\pm51$ signal events for the $D_s^+$ and $D^+$ respectively. The Dalitz
plot analyses are performed on events within $\sigma$ the nominal $D_s^+$ or
$D^+$ mass (Fig.~\ref{dalitz}).

\section{The decay amplitude}

The decay amplitude of the $D$ meson into the three-pion final state is written
as:
\begin{equation}
  A(D) = a_0 e^{i\delta_0} + F_1 + \sum_{i} a_i e^{i\delta_i} B(abc|r_i) ,
  \label{totampK}
\end{equation}
where the first term represents the direct non-resonant three-body amplitude
contribution, $F_1$ is the contribution of  $S$-wave states and the sum is over
the contributions from the intermediate two-body non-scalar resonances.
$B(abc|r_i)$ are the usual Breit-Wigner terms of the traditional isobar model,
whose explicit forms are given in \cite{e687_dpds}. $F_1$ is written in the
context of the \emph{K-matrix} approach which we will discuss shortly. The
coefficients and phases of the $A(D)$ amplitude are all relative to a free
parameter of the $F_1$ amplitude, $\beta_1$, whose modulus and phase are fixed
to 1 and 0 respectively (see below). The squared modulus of this amplitude
gives the probability density in the three pion Dalitz plot.

In general, the decay of a $D$ meson into three pions via a resonance involves
the production of a $(IJ)^{PC}$ state with an accompanying pion. It is believed
that, while the usual Breit-Wigner (BW) approximation is suitable for states
with $J>0$, since they are characterized by relatively narrow and isolated
resonances, the treatment of $S$-wave states requires a more general formalism
to account for non-trivial dynamics due to the presence of broad and
overlapping resonances \cite{pdg_rev,penn1}. For $J=0$, only states with even
isospin and positive $P$ and $C$ are allowed to strongly couple to
$\pi^+\pi^-$. We limit ourselves to isoscalar $S$-wave states, $(00)^{++}$,
since $I=2$  must involve at least two $q \bar q$ pairs and no four-quark
states with $I=2$ are known. At the mass scales relevant to this analysis, the
decay of a charm particle into a $(00)^{++}$ state with an accompanying pion
consists of five channels $l=1\dots5$ where $1=\pi\pi$, $2=K\bar{K}$,
$3=\mbox{multi-meson}$ states (four-pion state mainly at $\sqrt{s}<1.6\,$GeV)
$4=\eta\eta$ and $5=\eta\eta'$. The amplitude for the particular channel
$(00)^{++}_l\pi$ can be written in the context of the \emph{K-matrix} formalism
as
\begin{equation}
  F_l = (I-iK\rho)_{lj}^{-1}P_j \, ,
  \label{eq_f}
\end{equation}
where $I$ is the identity matrix, $K$ is the \emph{K-matrix} describing the
isoscalar $S$-wave scattering process, $\rho$ is the phase-space matrix for the
five channels, and $P$ is the ``initial" production vector into the five
channels. In this picture, the production process can be viewed as consisting
of an initial preparation of several states, which are then propagated by the
$(I-iK\rho)^{-1}$ term into the final one. Only the $F_1$ amplitude is present
in the isosinglet $S$-wave term since we are describing the dipion channel.

We require a reliable \emph{K-matrix} parametrization of $(00)^{++}$-wave
scattering. To our knowledge the only self-consistent description of the
$S$-wave isoscalar scattering is given by the \emph{K-matrix} representation of
Anisovich and Sarantsev in reference \cite{anisar1} obtained through a global
fit of the available scattering data from $\pi\pi$ threshold up to $1900\,$MeV.
Their \emph{K-matrix} parametrization is:
\begin{equation}
 K_{ij}^{00}(s) =
  \left\{
    \sum_\alpha \frac{g^{(\alpha)}_i g^{(\alpha)}_j}{m^2_{\alpha}-s}
    + f^{\scatt}_{ij}\frac{1\,\mathrm{GeV}^2 - s_0^{\scatt}}{s-s_0^{\scatt}}
  \right\}
  \frac{(s-s_A m^2_\pi/2)(1-s_{A0})}{(s-s_{A0})}.
  \label{eq_sarantsev}
\end{equation}
The $g^{(\alpha)}_i$ is the coupling constant of the \emph{K-matrix} pole
$m_\alpha$ to the $i$ meson channel; the parameters $f^{\scatt}_{ij}$ and
$s_0^{\scatt}$ describe a slowly varying part (which we will call SVP) of the
\emph{K-matrix} elements; the factor
$\frac{(s-s_Am^2_{\pi}/2)(1-s_{A0})}{(s-s_{A0})}$ is to suppress false
kinematical singularity in the physical region near the $\pi\pi$ threshold
(Adler zero). The parameter values used in this paper are listed in
Table~\ref{table_sara}, which was provided by the authors of reference
\cite{anisar1}. Note that \emph{K-matrix} representation is by definition real
and symmetric.
\begin{table}[htb]
  \caption{\emph{K-matrix} parameters. Masses and coupling constants
  are in GeV. Only the $i=1$ $f_{ij}$ terms are reported here since they are the
  only values relevant to the three-pion decay.}
 \label{table_sara}
$
\begin{array}{cccccc} \hline m_{\alpha} & g_{\pi \pi} & g_{K \bar K} &
g_{4 \pi} & g_{\eta \eta} & g_{\eta \eta' }\\
\hline
0.65100  & 0.24844&     -0.52523 &   \,\,0.00000&     -0.38878&     -0.36397 \\
1.20720  & 0.91779&  \,\,0.55427 &   \,\,0.00000&  \,\,0.38705&  \,\,0.29448 \\
1.56122  & 0.37024&  \,\,0.23591 &   \,\,0.62605&  \,\,0.18409&  \,\,0.18923 \\
1.21257  & 0.34501&  \,\,0.39642 &   \,\,0.97644&  \,\,0.19746&  \,\,0.00357 \\
1.81746  & 0.15770& -0.17915 &-0.90100& -0.00931&         \,0.20689 \\
\hline s_0^{\scatt} & f^{\scatt}_{11} & f^{\scatt}_{12} & f^{\scatt}_{13} &
f^{\scatt}_{14}
& f^{\scatt}_{15} \\
%\hline
-3.30564 & 0.26681 & 0.16583 & -0.19840 & 0.32808 & 0.31193\\
 \hline
s_A & s_{A0} & & & & \\
%\hline
1.0 & -0.2 & & & & \\
\hline
\end{array}
$
\end{table}
The \emph{K-matrix} values of Table~\ref{table_sara} generate a physical
\emph{T-matrix}, \mbox{$T=(I-i K\cdot\rho)^{-1}K$}, which describes the
scattering in the $(00)^{++}$-wave with five poles, whose masses, half-widths,
and couplings are listed in Table~\ref{table_scattering}.
%
%~it is better to report couplings as well
%\begin{table}[htb]
%\begin{tabular}{ccccc}
%\hline  $f_0(980)$\footnote{The f0(980) pole position is found on the second Riemann sheet;
% the closest physical region is the beginning of K $\bar K$ threshold.}
%& $f_0(1300)$& $f_0(1200-1600)$ &
%$f_0(1500$ & $f_0(1750)$ \\
%  (1.019,\,0.038)&  (1.306,\,0.170)&  (1.470,\,0.960)  &
%  (1.489,\,0.058)&1.746,\,0.160)\\
% \hline
%\end{tabular}
%\caption{ \emph{T-matrix} poles. The (m,-$\Gamma$/2) values are in GeV}
% \label{table_scattering}
%\end{table}

\begin{table}[htb]
  \caption{
    \emph{T-matrix} poles and $\pi^+\pi^-$ scattering coupling-constants;
    phases are in degrees.
    The $f_0(980)$ pole position is found by the authors of \cite{anisar1} on the second Riemann sheet,
    thus the
    closest physical region is the beginning of $K\bar K$ threshold.
  }
  \label{table_scattering}
\begin{tabular}{ccccc}
\hline  \emph{T-matrix} pole  & $g_{K \anti K}/g_{\pi \pi}$ & $g_{4
\pi}/g_{\pi\pi}
$ & $g_{\eta \eta}/g_{\pi \pi}$ & $g_{\eta \eta'}/g_{\pi \pi}$\\
 ( $m,\Gamma/2$ ) GeV  &                              &              & & \\
\hline $f_0(980)$
 &&&&\\
 (1.019,\,0.038)  &1.3970\,e$^{i83.4}$ &
0.3572\,e$^{i67.8}$ &1.1660\,e$^{i85.5}$ &0.9662\,e$^{i89.0}$ \\
\hline  $f_0(1300)$&&&&\\
(1.306,\,0.170)  & 0.2579\,e$^{-i16.6}$ &
2.1960\,e$^{-i178.7}$ &0.3504\,e$^{i23.2}$ &0.5547\,e$^{i16.2}$ \\
\hline  $f_0(1200-1600)$&&&&\\
(1.470,\,0.960)  & 1.1140\,e$^{-i0.4}$ &
2.2200\,e$^{-i6.8}$ &0.569\,e$^{i17.7}$ &0.2309\,e$^{i54.6}$ \\
\hline  $f_0(1500$)&&&&\\
 (1.488,\,0.058)   & 0.5460\,e$^{-i1.8}$ &
1.9790\,e$^{i85.3}$ &
 0.4083\,e$^{i37.9}$ &  0.4692 e$^{i74.6}$\\
\hline  $f_0(1750)$&&&&\\
(1.746,\,0.160)  & 0.1338\,e$^{i32.6}$ & 1.3680\,e$^{i134.8}$ &
 0.2979 e$^{i25.1}$ &  0.5843 e$^{-i0.5}$\\
\hline%
\end{tabular}
\end{table}

The $f_0$ series reported in Table~\ref{table_scattering} differs somewhat from
that reported by the PDG \cite{PDG} group. In addition to the $f_0(980)$ and $f_0(1500)$
poles which also appear in the PDG classification, three other poles are
present, $f_0(1300)$, $f_0(1200-1600)$ and $f_0(1750)$, in contrast with only
two poles listed by the PDG, $f_0(1370)$ and $f_0(1710)$. The five-pole series
used here is able to consistently reproduce the available $S$-wave isoscalar
data in the energy range relevant for this analysis. The decay amplitude for
the $D$ meson into the three-pion final state, where $\pi^+\pi^-$ are in a
$(IJ^{PC}=00^{++}$)-wave is then
\begin{eqnarray}
  F_1 &=&
  (I-iK\rho)^{-1}_{1j}
  \left\{
    \sum_\alpha \frac{\beta_{\alpha}g_{j}^{(\alpha)}}{m^2_{\alpha}-s}
    + f_{1j}^{\mathrm{prod}}\frac{1\,\mathrm{GeV}^2 - s_0^{\mathrm{prod}}}{s-s_0^{\mathrm{prod}}}
  \right\}
  \nonumber
\\
  &&\hspace{16em} \null \times
  \frac{(s-s_A m^2_{\pi}/2)(1-s_{A0})}{(s-s_{A0})}
  \, ,
  \label{eq_kmat}
\end{eqnarray}
where $\beta_{\alpha}$ is the coupling to the $m_{\alpha}$  pole in the
`initial' production process, $f_{1j}^{\mathrm{prod}}$ and
$s_0^{\mathrm{prod}}$ are the \emph{P-vector} SVP parameters. $\beta_{\alpha}$
and $f_{1j}^{\mathrm{prod}}$ are in general complex numbers \cite{chung}. The phase space
matrix elements for the two pseudoscalar-particle states are:
\begin{eqnarray}
  \rho_i(s) &=&
  \sqrt{\left(1-\frac{(m_{1i}+m_{2i})^2}{s}\right)
        \left(1-\frac{(m_{1i}-m_{2i})^2}{s}\right)}
  \nonumber
\\
  &&\hspace{16em}
  (i=\pi\pi, K\bar K,\eta\eta, \eta\eta').
  \label{rho}
\end{eqnarray}
The normalization is such that $\rho_i\to1$ as $s\to\infty$. The expression for
the multi-meson state phase space can be found in reference \cite{anisar1}.

We note that the \emph{P-vector} poles have to be the same as those of the
\emph{K-matrix} in order to cancel out infinities in the final amplitude as
each pole is realized. The \emph{P-vector} SVP parametrization is chosen in
complete analogy with that used for the \emph{K-matrix}. The need for the
Adler-zero term, not \emph{a-priori} required in the \emph{P-vector}, will be
investigated by studying its effect on the quality of the fit to our data. The
\emph{K-matrix} parameters are fixed to the values of Table~\ref{table_sara} in
our Dalitz plot fits.  The free parameters are the \emph{P-vector} parameters
($\beta_{\alpha}$, $f^{\mathrm{prod}}_{1j}$ and $s_0^{\mathrm{prod}}$), and the
coefficients and phases of Eq.~\ref{totampK} ($a_0,a_i$ and
$\delta_0,\delta_i$). All amplitudes are referenced to $\beta_1$ which is fixed
at 1. The \emph{P-vector} Adler-zero parameters have been chosen to be
identical to those of the  \emph{K-matrix}, $s_A$ and $s_{A0}$. Because of the
limited allowed range for these parameters, we do not expect our results to
critically depend on this particular choice.

\section{The likelihood function and fitting procedure}

The probability density function is corrected for geometrical acceptance and
reconstruction efficiency. We find that finite-mass resolution effects are
negligible. The shape of the background in the signal region is parametrized
through a polynomial fit to the Dalitz plot of mass sidebands\footnote{ For
this analysis the sideband between the two signal peaks begins at $3\,\sigma$
from the $D^+$ peak and ends at $3\,\sigma$ from the $D_s^+$ peak where the
$\sigma$'s are the r.m.s.\ widths of the two measured mass peaks. The left
sideband for $D^+$ covers the $-5\,\sigma$ to $-3\,\sigma$ region from the
$D^+$ peak, while the right sideband for $D_s^+$, the $3\,\sigma$ through
$6\,\sigma$ region from the $D_s^+$ mass peak.}. The number of background
events expected in the signal region is estimated through fits to the
$\pi^+\pi^-\pi^+$ mass spectrum. All background parameters are included as
additional fit parameters and tied to the results of the sideband fits through
the inclusion in the likelihood of a $\chi^2$ penalty term derived from the
covariance matrix of the sideband fit. The contamination in the $D^+$ left
sideband from $D^+\to K^-\pi^+\pi^+$, where $K^-$ is misidentified as $\pi^-$,
is reduced to a negligible level (3.5\,\% of the total events in the sideband)
using the tight \v{C}erenkov cut. Background from the decay,
$D_s^+\to\eta'\pi^+$ with $\eta'\to\rho\gamma$ and $\rho\to\pi^+\pi^-$, is
expected in the $D^+$ signal and sideband regions. It is included by adding a
$\rho$-BW component in the background parametrization. The $D^+$ and $D_s^+$
samples are fitted with likelihood functions $\mathcal{L}$ consisting of signal
and background probability densities. Checks for fitting procedure are made
using Monte Carlo techniques and all biases are found to be small compared to
the statistical errors. The systematic errors on our results are evaluated by
comparing their values in disjoint samples corresponding to different
experimental running conditions, different kinematical regions, such as low
versus high $D$ momenta, and particle versus anti-particle. A split sample
systematic error was added in quadrature to the existing statistical error to
make the split sample estimates consistent to within $1\,\sigma$ if necessary.
The assumption that the shape of the background in the sideband is a good
representation of the background in the signal region could potentially
constitute another source of systematic error. We study this effect by varying
the polynomial function degree and adding/removing the Breit-Wigner terms,
which are introduced to take into account any feed-through from resonances in
the background, and computing the r.m.s.\ of the different results, which is
added in quadrature to form the total experimental systematic error.
%jew

\section{\boldmath Results for the $D^+_s\to\pi^+\pi^-\pi^+$ decay}

We recall that the physical parameters of our fit are
 \emph{P-vector} parameters:
$\beta_{\alpha}$, $f^{\mathrm{prod}}_{1j}$, along with $s_0^{\mathrm{prod}}$,
and the coefficients and phases of Eq.~\ref{totampK}, $a_0,a_i$ and
$\delta_0,\delta_i$. The \emph{K-matrix} parameters are fixed to the values
given in Table~\ref{table_sara}. The general procedure, adopted for all the
fits reported here, consists of several successive steps in order to eliminate
contributions whose effects on our fit are marginal. We initially consider all
the well established, non-scalar resonances decaying to $\pi^+\pi^-$ with a
sizeable branching ratio. Contributions are removed if their amplitude
coefficients, $a_i$ of Eq.~\ref{totampK}, are less than $2\,\sigma$ significant
\emph{and} the fit confidence level increases due to the decreased number of
degrees of freedom in the fit. The \emph{P-vector} initial form includes the
complete set of \emph{K-matrix} poles and slowly varying function (SVP) as
given in reference \cite{anisar1}; $\beta_{\alpha}$ as well as the
$f^{\mathrm{prod}}_{1j}$ terms of Eq.~\ref{eq_kmat} are removed with the same
criteria. The fit confidence levels (C.L.) are evaluated with a $\chi^2$
estimator over a Dalitz plot with bin size adaptively chosen to maintain a
minimum number of events in each bin. Once the minimal set of parameters is
reached, addition of each single contribution previously eliminated is
reinstated to verify that the C.L. does not improve.

Table~\ref{table_ds_pvector} shows the \emph{P-vector} composition from our
final fit results on the $D_s^+\to\pi^+\pi^-\pi^+$ Dalitz plot.
%\begin{table}
%\centering
%\begin{tabular}{ccc}
%\hline
%\emph{P-vector} component & coefficient & phase (deg)  \\
%\hline
%$\beta_1  $   &  1 (fixed)          & 0 (fixed)           \\
%$\beta_2  $   &  0.728$\pm$ 0.076   & 207.0 $\pm$5.9     \\
%$\beta_3  $   &  0.292$\pm$ 0.076   & 85.7 $\pm$19.3     \\
%$\beta_4  $   &  0.436$\pm$ 0.092   & 119.1 $\pm$19.9     \\
%$f^{\mathrm{prod}}_{11} $   &  0.552$\pm$ 0.098   & 202.3 $\pm$9.4      \\
%$f^{\mathrm{prod}}_{13} $   &  0.389$\pm$ 0.077   & 140.3 $\pm$16.7     \\
%$f^{\mathrm{prod}}_{14} $   &  0.509$\pm$ 0.155   & 106.6 $\pm$24.4      \\
%$f^{\mathrm{prod}}_{15} $   &  1.046$\pm$ 0.292   & 236.5 $\pm$13.7      \\
%\hline
%\end{tabular}
%\caption{ \emph{P-vector} coefficients and phases returned by the $D_s^+$ fit}
% \label{table_ds_pvector}
%\end{table}
%
\begin{table}[htb]
  \caption{ \emph{P-vector} parameters of the $D_s^+$ fit.}
  \label{table_ds_pvector}
\begin{tabular}{ccc}
\hline
\emph{P-vector} parameters & modulus & phase (deg)  \\
\hline
$\beta_1$   &  1 (fixed)           & 0 (fixed)           \\
$\beta_2$   &  $0.974 \pm 0.102$   & $ 207.0 \pm 5.9 $    \\
$\beta_3$   &  $0.755 \pm 0.196$   & $ \,\,\,85.7  \pm 19.3$     \\
$\beta_4$   &  $1.760 \pm 0.371$   & $ 119.1 \pm 19.9$     \\
$f^{\mathrm{prod}}_{11} $   &  $0.863 \pm 0.153$   & $202.3 \pm 9.4$      \\
$f^{\mathrm{prod}}_{13} $   &  $3.768 \pm 0.746$   & $140.3 \pm 16.7$     \\
$f^{\mathrm{prod}}_{14} $   &  $2.179 \pm 0.663$   & $106.6 \pm 24.4$      \\
$f^{\mathrm{prod}}_{15} $   &  $2.452 \pm 0.684$   & $236.5 \pm 13.7$      \\
\hline
\end{tabular}
\end{table}
The fifth \emph{K-matrix} pole and the second SVP contribution were eliminated.
The inclusion of an Adler zero term did not improve our fit quality and was
removed. The quoted results were obtained with $s^{\mathrm{prod}}_0=-3$
GeV$^2$, but they were insensitive to any choice in the range $-5\,$GeV$^2\le
s^{\mathrm{prod}}_0 \le -1\,$GeV$^2$ -- typical parameter values for the SVP.

\begin{figure}[htb]
\centering
 \subfigure
  {
  \includegraphics[width=0.9\textwidth]{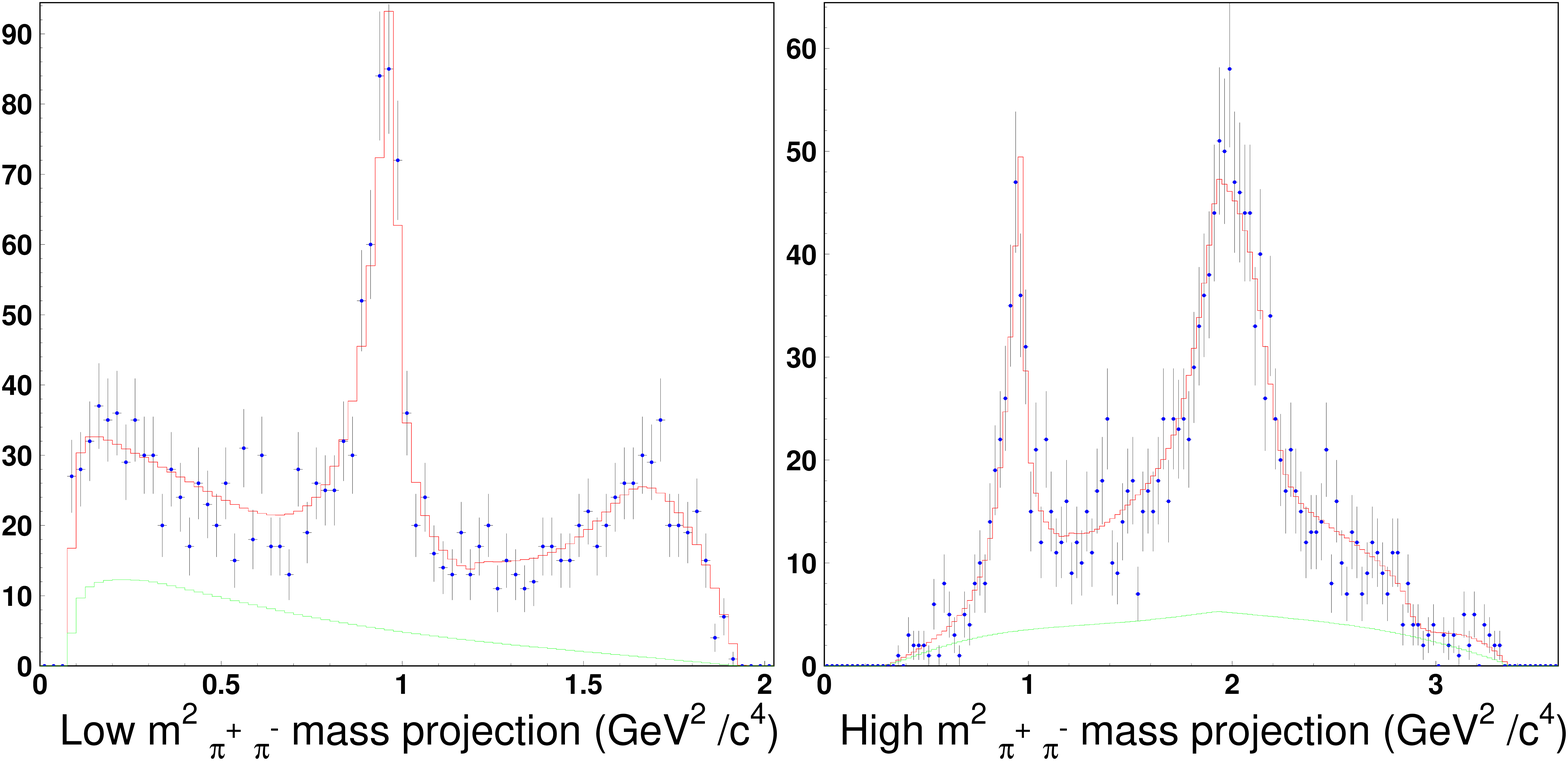}
  }
 \caption{$D_s^+$ Dalitz-plot projections with our final fit
superimposed. The background shape under the signal is also shown.}
\label{ds_proj_Kmatrix}
\end{figure}

\begin{figure}[htb]
\centering
\subfigure
  {
  \includegraphics[width=0.5\textwidth]{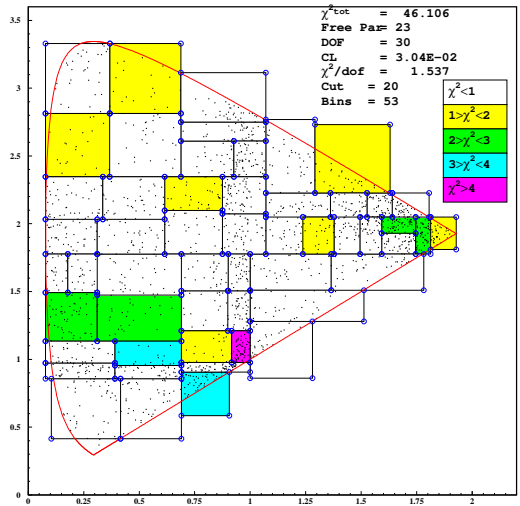}
  }
 \caption{$D_s^+$ adaptive binning Dalitz-plot for the \emph{K-matrix} model.}
 \label{ds_adapt_km}
\end{figure}

The resulting fit fractions \footnote{ The quoted fit fractions are defined as
the ratio between the intensity for a single amplitude integrated over the
Dalitz plot and that of the total amplitude with all the modes and
interferences present.}, phases and amplitude coefficients are quoted in
Table~\ref{table_ds_kmatrix}. We note that both the three-body non-resonant and
$\rho^0(770)\pi^+$ components were not required by the fit. We represent the
entire $S$-wave contribution by a single fit fraction since, as previously
discussed, one cannot distinguish the different resonance or SVP $S$-wave
contributions on the real axis. The couplings to \emph{T-matrix} physical
poles, reported in Table~\ref{table_fvect_Ds}, are computed by continuing the
amplitude $F_1(s)$ into the complex s-plane to the position of the poles and
evaluating the pole residues \footnote{The coupling to the $f_0(1200-1600)$
pole has to be looked at with a certain caution because of the intrinsic
limitations of the approximation used for the 4-body phase-space when
extrapolated very deeply into the complex plane.}. The $D_s^+$ Dalitz
projections of our data are shown in Fig.~\ref{ds_proj_Kmatrix} superimposed
with our final fit projections. Figure~\ref{ds_adapt_km} shows the
corresponding adaptive binning scheme used to obtain the fit confidence level.
\begin{table}[htb]
  \caption{ Fit results from the \emph{K-matrix} model for $D_s^+$.}
  \label{table_ds_kmatrix}
\begin{tabular}{cccc}
\hline
decay channel  & fit fraction (\%) & phase (deg) & amplitude coefficient \\
\hline ($S$-wave)\,$\pi^+$     &  $87.04 \pm 5.60 \pm 4.17$ & 0 (fixed)
     & 1 (fixed)  \\
$f_2(1270)\,\pi^+$    &  $9.74 \pm 4.49 \pm 2.63$&  $168.0 \pm 18.7 \pm 2.5$
 & $0.165 \pm 0.033\pm 0.032$ \\
$\rho^0(1450)\,\pi^+$ &  $6.56 \pm 3.43 \pm 3.31$ &  $234.9 \pm 19.5\pm 13.3$
& $0.136\pm 0.030 \pm 0.035$  \\
\hline Fit C.L & 3.0\,\% & &\\
\hline
\end{tabular}
\end{table}
\begin{table}[htb]
  \caption{$D_s^+\to\pi^+\pi^-\pi^+$ production coupling constants for the five
  \emph{T-matrix} poles, which comprise the $S$-wave component in
  Table~\ref{table_ds_kmatrix}, referenced to the $f_0(980)$. Phases
  are in degrees.}
  \label{table_fvect_Ds}
\begin{tabular}{ccc}
\hline
\emph{T-matrix} pole  &  ($m,\Gamma/2$) (GeV)& $D_s^+$ (relative) coupling constant\\
\hline
$f_0(980)$ & (1.019,\,0.038) & 1 e$^{i 0}$ (fixed) \\
$f_0(1300)$&(1.306,\,0.170)  & $(0.43 \pm 0.04)$ e$^{i(-163.8 \pm 4.9)}$ \\
$f_0(1200-1600)$&(1.470,\,0.960) & $(4.90 \pm 0.08)$ e$^{i(\,\,\,80.9 \pm
1.06)}$\\
$f_0(1500)$&(1.488,\,0.058) & $(0.51 \pm 0.02)$ e$^{i(\,\,\,83.1 \pm 3.03)}$\\
$f_0(1750)$&(1.746,\,0.160) & $(0.82 \pm0.02)$  e$^{i(-127.9 \pm 2.25)}$ \\
\hline
 \end{tabular}
\end{table}

\section{\boldmath Results for the $D^+\to\pi^+\pi^-\pi^+$ decay}

The $D^+\to\pi^+\pi^-\pi^+$ Dalitz plot shows an excess of events at low
$\pi^+\pi^-$ mass, which cannot be  explained in the context of the simple
isobar model with the usual mixture of well established resonances along with a
constant, non-resonant amplitude. A new scalar resonance, the $\sigma(600)$,
has been previously proposed \cite{e791_dp} to describe this excess. However we
know that complex structure can be generated by the interplay among the
$S$-wave resonances and the underlying non-resonant $S$-wave component that
cannot be properly described in the context of a simple isobar model. It is
therefore interesting to study this channel with the present formalism, which
embeds all our experimental knowledge about the $S$-wave $\pi^+\pi^-$
scattering dynamics.

With the same procedure based on statistical significance and fit confidence level
used in the $D_s^+$ analysis, we obtained the final set of \emph{P-vector} parameters
that is reported in Table~\ref{table_dp_pvector}.
%
%\begin{table}
%\centering
%\begin{tabular}{ccc}
%\hline
%\emph{P-vector} component & coefficient & phase (deg)  \\
%\hline
%$\beta_1 $   &  1 (fixed)          & 0(fixed)            \\
%$\beta_2 $   &  1.281$\pm$ 0.391   & 104.8 $\pm$17.1     \\
%$\beta_3 $   &  1.103$\pm$ 0.253   & 170.7 $\pm$13.7     \\
%$\beta_4 $   &  0.754$\pm$ 0.201   & 145.2 $\pm$21.4     \\
%$f^{\mathrm{prod}}_{11} $   &  2.466$\pm$ 0.629   & 139.1 $\pm$23.0      \\
%\hline
%\end{tabular}
%\caption{ \emph{P-vector} coefficients and phases returned by the $D^+$ fit}
%\label{table_dp_pvector}
%\end{table}
%
\begin{table}[htb]
 \caption{\emph{P-vector} parameters of the $D^+$ fit.}
 \label{table_dp_pvector}
\begin{tabular}{ccc}
\hline
\emph{P-vector} parameters & modulus & phase (deg)  \\
\hline
$\beta_1 $   &  1 (fixed)          & 0 (fixed)            \\
$\beta_2 $   &  $2.471\pm 0.431$   & $ 82.5 \pm  10.3$     \\
$\beta_3 $   &  $1.084\pm 0.386$   & $102.8 \pm  23.5$     \\
$f^{\mathrm{prod}}_{11} $   &  $2.565\pm 0.737$   & $155.4 \pm 18.3 $ \\
$f^{\mathrm{prod}}_{12} $   &  $6.312\pm 0.967$   & $-160.0 \pm 8.7 $ \\
\hline
\end{tabular}
\end{table}
The last two poles and the last three SVP terms were eliminated. The
$s^{\mathrm{prod}}_0$ value is measured to be
$s^{\mathrm{prod}}_0=(-1.0^{+0.4}_{-5.5})\,$GeV$^2$ . The fit did not require an
Adler-zero term.

\begin{figure}[htb]
  \centering
  \subfigure
  {
  \includegraphics[width=0.9\textwidth]{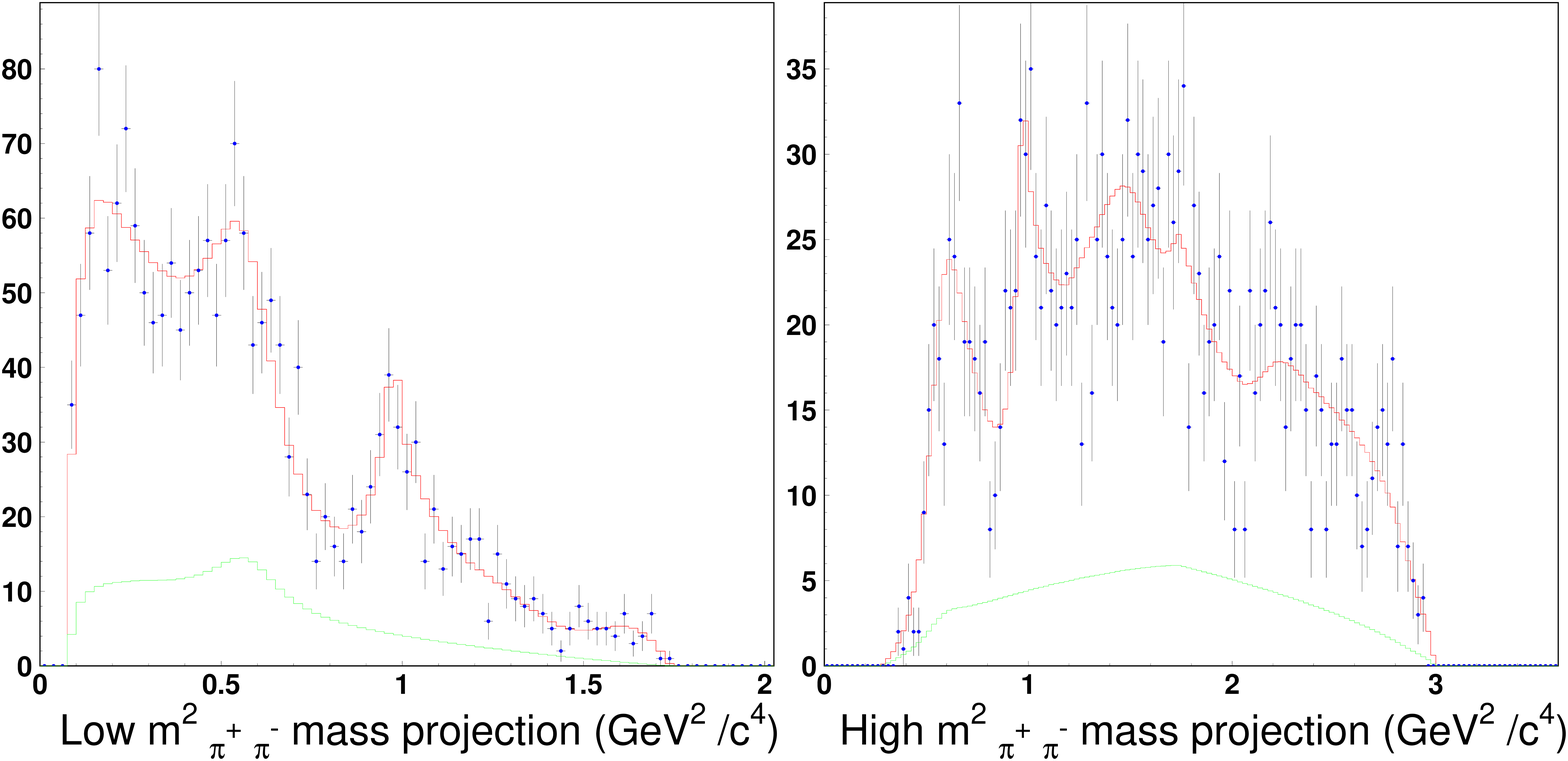}
  }
  \caption{$D^+$ Dalitz-plot projections  with our final fit
  superimposed. The background shape under the signal is also shown.}
  \label{dp_proj_Kmatrix}
\end{figure}

\begin{figure}[htb]
  \centering
  \subfigure
  {
  \includegraphics[width=0.5\textwidth]
  {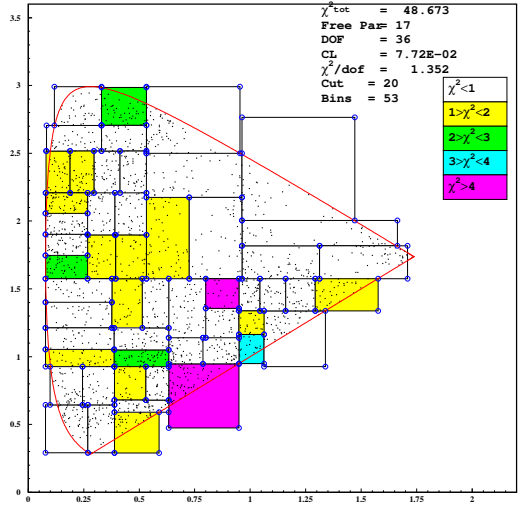}
  }
  \caption{$D^+$ adaptive binning Dalitz-plot for the \emph{K-matrix} fit.}
  \label{dp_adapt_km}
\end{figure}

Beside the $S$-wave component, the decay appears to be dominated by the
$\rho^0(770)$ plus a $f_2(1270)$ component. The $\rho^0(1450)$ was always found
to have less than $2\,\sigma$ significance and was therefore dropped from the
final fit. In analogy with the $D_s^+$, the direct three-body non-resonant
component was not necessary since the SVP of the $S$-wave could reproduce the
entire non-resonant portion of the Dalitz plot. The complete fit results are
reported in Table~\ref{table_dp_kmatrix}. The resulting production coupling
constants are reported in Table~\ref{table_fvect_Dp}.
\begin{table}[htb]
 \caption{ Fit results from the \emph{K-matrix} model fit for $D^+$.}
 \label{table_dp_kmatrix}
\begin{tabular}{cccc}
\hline
decay channel  & fit fraction (\%) & phase (deg) & amplitude coefficient \\
\hline
($S$-wave)\,$\pi^+$   &  $56.00 \pm 3.24 \pm 2.08$  & 0 (fixed)            &  1 (fixed)    \\
$f_2(1270)\,\pi^+$    &  $11.74 \pm 1.90 \pm 0.23$   & $-47.5 \pm 18.7\pm 11.7$ &
$1.147\pm 0.291 \pm0.047$ \\
$\rho^0(770)\,\pi^+$  &  $30.82 \pm 3.14 \pm 2.29$  &$-139.4 \pm  16.5 \pm 9.9$  & $1.858 \pm 0.505 \pm0.033$
  \\
\hline
 Fit C.L. & 7.7\,\% & & \\
\hline
\end{tabular}
\end{table}
\begin{table}[htb]
 \caption{$D^+\to\pi^+\pi^-\pi^+$ production coupling constants  for the five
 \emph{T-matrix} poles, which comprise the $S$-wave component
 in Table~\ref{table_dp_kmatrix}, referenced to the $f_0(980)$. Phases are in degrees.}
 \label{table_fvect_Dp}
\begin{tabular}{ccc}
\hline \emph{T-matrix} pole  & ($m,\Gamma/2$) (GeV)& $D^+$ (relative)
coupling constant\\
\hline
$f_0(980)$ & (1.019,\,0.038)      & 1 e$^{i0}$ (fixed)\\
$f_0(1300)$ &(1.306,\,0.170)      &  $(0.67 \pm 0.03)$ e$^{i(-67.9 \pm 3.0)}$ \\
$f_0(1200-1600)$ &(1.470,\,0.960) &  $(1.70 \pm 0.17)$ e$^{i(-125.5\pm 1.7)}$\\
$f_0(1500)$ &(1.488,\,0.058)      &  $(0.63 \pm 0.02)$ e$^{i(-142.2\pm 2.2)}$\\
$f_0(1750)$ &(1.746,\,0.160)      &  $(0.36 \pm 0.02)$ e$^{i(-135.0 \pm 2.9)}$\\
 \hline
 \end{tabular}
\end{table}
The $D^+$ Dalitz projections are shown in Fig.~\ref{dp_proj_Kmatrix} and the
corresponding adaptive binning scheme is shown in Fig.~\ref{dp_adapt_km}.
%
%At this level of statistics the strictly necessary \emph{P-vector} parameters
%are those reported in Table~\ref{table_dp_pvector}. However it is worth
%showing here the quality of the fit performed with the complete set of
%\emph{P-vector} parameters; the Dalitz plot projections, drawn in
%Fig.~\ref{dp_all_free}, correspond to a fit C.L. of 6.4 \%.
%
%\begin{figure}[htb]
%\centering
%  %%\subfigure
%  {
%  \includegraphics[width=0.9\textwidth]{kmatrix_dp.ps}
%  }
%\caption{$D^+$  Dalitz-plot projections with the complete set of
%\emph{P-vector} parameters.}
% \label{dp_all_free}
%\end{figure}
%jew
The most interesting feature of these results is the fact that the better
treatment of the $S$-wave contribution provided by the \emph{K-matrix} model
can reproduce the low-mass $\pi^+\pi^-$ structure of the $D^+$ Dalitz plot.
This suggests that any $\sigma$-like object in the $D$ decay should be
consistent with the same $\sigma$-like object measured in the $\pi^+\pi^-$
scattering. We believe that additional studies with higher statistics will be
required to completely understand the $\sigma$ puzzle.

\section{\boldmath $D_s^+$ and $D^+$ final results}

The \emph{K-matrix} parameters used in this analysis correspond to the best
solution provided by the authors of reference \cite{anisar1}. Several solutions
with slightly different parametrizations for the 4$\pi$ phase-space and for the
\emph{K-matrix} background terms were presented in the same paper. We evaluate
the systematic error due to solution choice by computing the r.m.s.\ of the fit
fractions and phases obtained using the different solutions. The final results,
including this last systematic error, are presented in Table~\ref{table_sol}.
\begin{table}[htb]
  \caption{Final results on $D_s^+$ and $D^+\to\pi^+\pi^-\pi^+$  fit
  fractions and phases. Beside the first reported error, which is statistical,
  two systematic errors are quoted. The first one is from the measurement
  systematics and the second one is due to the particular solution chosen for the
  \emph{K-matrix} poles and backgrounds.}
  \label{table_sol}
\begin{tabular}{ccc}
\hline
 \multicolumn{3}{c}{$D_s^+$} \\
\hline
decay channel & fit fraction (\%) & phase (deg)\\
($S$-wave)\,$\pi^+$  & $87.04 \pm 5.60 \pm 4.17 \pm 1.34$ & 0 (fixed)\\
 $f_2(1270)\,\pi^+$  &  $9.74 \pm 4.49 \pm 2.63 \pm 1.32$ & $168.0 \pm 18.7 \pm2.5 \pm 21.7$ \\
 $\rho^0(1450)\pi^+$   &  $6.56 \pm 3.43 \pm 3.31 \pm 2.90$ & $234.9 \pm
 19.5 \pm 13.3 \pm 24.9$ \\
 \hline
 \multicolumn{3}{c}{$D^+$} \\
 \hline
decay channel & fit fraction (\%) & phase (deg)\\
($S$-wave)\,$\pi^+$    & $56.00 \pm 3.24 \pm 2.08 \pm 0.50$ & 0 (fixed)\\
 $f_2(1270)\,\pi^+$    & $11.74 \pm 1.90 \pm 0.23 \pm 0.18$   & $-47.5\pm 18.7 \pm
  11.7 \pm 5.3 $\\
 $\rho^0(770)\pi^+$   &    $30.82 \pm 3.14 \pm 2.29 \pm 0.17$   &$-139.4 \pm 16.5
 \pm 9.9 \pm 5.0$ \\
 \hline
 \end{tabular}
\end{table}

\section{Conclusions}

The \emph{K-matrix} formalism has been applied for the first time to the charm
sector in our Dalitz plot analyses of the $D_s^+$ and $D^+\to\pi^+\pi^-\pi^+$
final states. The results are extremely encouraging since the same
\emph{K-matrix} description gives a coherent picture of both  two-body
scattering measurements in light-quark experiments \emph{as well as} charm
meson decay. This result was not obvious beforehand. Furthermore, the same
model is able to reproduce features of the $D^+\to\pi^+\pi^-\pi^+$ Dalitz plot
that otherwise would require an \emph{ad hoc} $\sigma$ resonance. In addition,
the non-resonant component of each decay seems to be described by known
two-body $S$-wave dynamics without the need to include constant amplitude
contributions.

\begin{figure}[htb]
\centering
  \includegraphics[width=1.\textwidth]{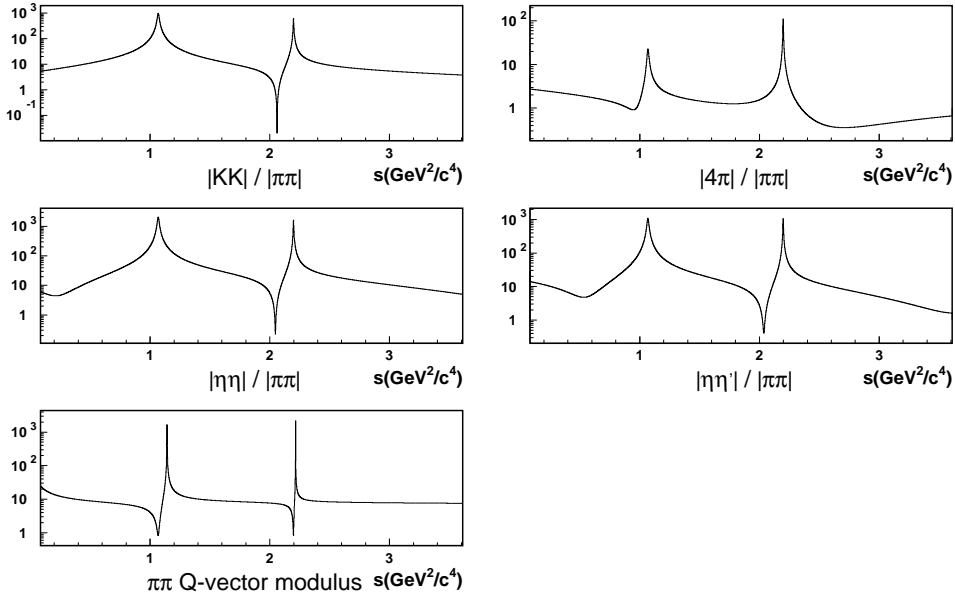}
\caption{Measured $D_s^+$ \emph{Q-vector} components: the first four plots are
the ratios of moduli of the \emph{Q-vector} amplitudes with respect to the
$\pi\pi$ modulus; the last plot is the normalizing $\pi\pi$ modulus.}
 \label{qvector_ds}
\end{figure}

\begin{figure}[htb]
\centering
 \subfigure
  {
  \includegraphics[width=1.\textwidth]{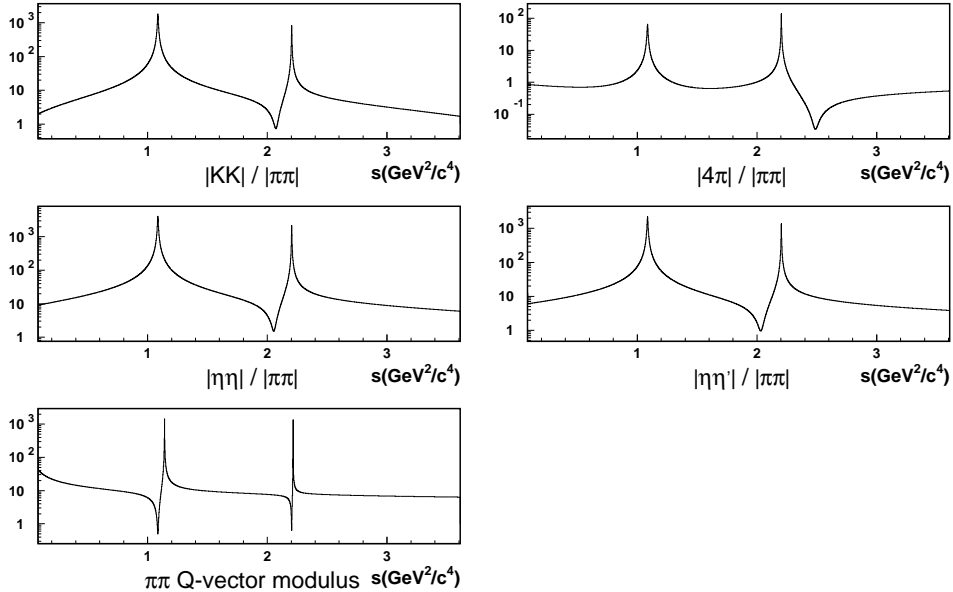}
  }
\caption{Measured $D^+$ \emph{Q-vector} components: the first four plots are
the ratios of moduli of the \emph{Q-vector} amplitudes with respect to the
$\pi\pi$ modulus; the last plot is the normalizing $\pi\pi$ modulus.}
 \label{qvector_dp}
\end{figure}

The \emph{K-matrix} treatment of the $S$-wave component of the decay amplitude
allows for a direct interpretation of the decay mechanism in terms of the five
virtual channels considered:  $\pi\pi$, $K\bar K$, $\eta\eta$, $\eta\eta '$ and
$4\,\pi$. By inserting $KK^{-1}$ in the decay amplitude, $F$,
\begin{equation}
  F = (I-iK\rho)^{-1}P=(I-iK\rho)^{-1}KK^{-1}P=TK^{-1}P=TQ
  \label{q-vect}
\end{equation}
we can view the decay as consisting of an initial production of the five
virtual states which then scatter via the physical $T$ into the final state.
The \emph{Q-vector} contains the production amplitude of each virtual channel
in the decay. Figure~\ref{qvector_ds} shows the ratio of the moduli of the
\emph{Q-vector} amplitudes with respect to the $\pi\pi$ modulus for the $D_s^+$
$S$-wave. The last plot in Fig.~\ref{qvector_ds} represents the normalizing
$\pi\pi$ modulus. The two peaks of the ratios correspond to the two dips of the
$\pi\pi$ normalizing modulus, while the two peaks due to the \emph{K-matrix}
singularities, visible in the normalization plot, cancel out in the ratios.
Figure~\ref{qvector_dp} shows the analogous plots for the $D^+$ $S$-wave decay.
The resulting picture, for both $D_s^+$ and $D^+$ decay, is that the $S$-wave
decay is dominated by an initial production of $\eta\eta$, $\eta\eta'$ and
$K\bar K$ states. Dipion production is always much smaller. This suggests that
in both cases the $S$-wave decay amplitude primarily arises from a $s\bar s$
contribution such as that produced by the Cabibbo favoured weak diagram for the
$D_s^+$ and one of the two possible singly Cabibbo suppressed diagrams for the
$D^+$. For the $D^+$, the $s\bar s$ contribution competes with a $d\bar d$
contribution. That the $f_0(980)$ appears as a peak in the $\pi\pi$ mass
distribution in $D^+$ decay, as it does in $D_s$ decay, shows that for the
$S$-wave component the $s{\bar s}$ contribution dominates~\cite{penn1}.
Comparing the relative $S$-wave fit fractions that we observe for $D_s^+$ and
$D^+$ reinforces this picture. The  $S$-wave decay fraction for the $D_s^+$
(87\,\%) is larger than that for the $D^+$ (56\,\%). Rather than coupling to an
$S$-wave dipion, the $d\bar d$ piece prefers to couple to a vector state like
$\rho^0(770)$ that alone accounts for $\sim30\,\%$ of $D^+$ decay.

This interpretation also bears on the role of the annihilation diagram in the
$D_s^+\to\pi^+\pi^-\pi^+$ decay. We believe that Fig.~\ref{qvector_ds} suggests
that the $S$-wave annihilation contribution is negligible over much of the
dipion mass spectrum. It might be interesting to search for annihilation
contributions in higher spin channels, such as $\rho^0(1450)\pi$ and
$f_2(1270)\pi$.

\section{Acknowledgments}
We are particularly indebted to Prof. M.~R.~Pennington, for his patience in
guiding us through the fascinating \emph{K-matrix} world and for his frequent
advice in formalizing our problem. This work would have not be possible without
the invaluable help and assistance by Prof. V.~V.~Anisovich and Prof.
A.~V.~Sarantsev, who provided us with \emph{K-matrix} input numbers and even
crucial pieces of code. Their expertise was vital to us and certainly
accelerated our work. We wish to acknowledge the assistance of the staffs of
Fermi National Accelerator Laboratory, the INFN of Italy, and the physics
departments of the collaborating institutions. This research was supported in
part by the US National Science Fundation, the US Department of Energy, the
Italian Istituto Nazionale di Fisica Nucleare and Ministero dell'Istruzione
dell'Universit\`a e della Ricerca, the Brazilian Conselho Nacional de
Desenvolvimento Cient{\'\i}fico e Tecnol\'ogico, CONACyT-M\'exico, the Korean
Ministry of Education, and the Korean Science and Engineering Foundation.


\begin{thebibliography}{99}

\bibitem{pdg_rev}
S.~Spanier and N.~A.~T\"ornqvist, Scalar Mesons (rev.),
  Particle Data Group, Phys.~Rev.~D66 (2002) 010001-450.

\bibitem{penn1}
M.~R.~Pennington,
%``In Search Of Hadrons Beyond The Quark Model,"
  \emph{Proc. of Oxford Conf. in honour of R.~H.~Dalitz}, Oxford, July, 1990, Ed.
  by I.~J.~R. Aitchison, \emph{et al.}, (World Scientific) pp.~66--107; \emph{Proc. of
  Workshop on Hadron Spectroscopy} (WHS~99), Rome, March 1999, Ed.\ by
  T.~Bressani \emph{et al.}, (INFN, Frascati).

\bibitem{wigner}
E.~P.~Wigner, Phys.~Rev.~70 (1946) 15.

\bibitem{chung}
S.~U.~Chung \emph{et al.},
  Ann.~Physik~4 (1995) 404.

\bibitem{aitch}
I.~J.~R.~Aitchison, Nucl.~Phys. A189, (1972) 417.

\bibitem{penn2}
K.~L.~Au, D.~Morgan, and M.~R.~Pennington, Phys.~Rev.~D35 (1987) 1633.

\bibitem{anisar1}
V.~V.~Anisovich and A.~V.~Sarantsev,
  Eur.~Phys.~J. A16 (2003) 229.

\bibitem{nimcer}
J.~M.~Link \emph{et al.}, Nucl.~Instr.~Meth. A484 (2002) 270.

\bibitem{e687_dpds}
P.~L.~Frabetti \emph{et al.}, Phys.~Lett. B407 (1997) 79.

\bibitem{PDG}
  Particle Data Group, Phys.~Rev.~D66 (2002) 010001.

\bibitem{e791_dp}
E.~M.~Aitala \emph{et al.}, Phys.~Rev.~Lett. 86 (2001) 770.

\bibitem{cahn-land}
R.~N.~Chan and P.~V.~Landshoff, Nucl.~Phys. B266 (1986) 451.

\end{thebibliography}
\end{document}